\documentclass[twocolumn,10pt]{IEEEtran}

\usepackage{amsmath}
\usepackage{amsfonts}
\usepackage{epsfig}
\usepackage{amssymb}
\usepackage{cite}
\usepackage{color,soul}
\usepackage{subfigure}
\usepackage{multirow}
\usepackage{rotating}
\usepackage{graphicx}
\usepackage{tabularx}
\usepackage{array}
\usepackage{graphicx,dblfloatfix}
\usepackage{blindtext}
\usepackage{ragged2e}
\usepackage{xcolor,colortbl}
\definecolor{Gray}{gray}{0.85}

\usepackage{amsthm,amssymb,amsmath,bm}
\usepackage{fancyhdr}
\usepackage[subfigure]{tocloft}
\usepackage{ptext}
\usepackage[font={small}]{caption}
\usepackage{algorithmic}
\usepackage[Algorithm]{algorithm}
\usepackage{multicol}
\usepackage[subfigure]{tocloft}
\usepackage[font={small}]{caption}

\begin{document}
 \title{Robust Power Allocation in Covert Communication: Imperfect CDI}
     \author{\IEEEauthorblockN{Moslem Forouzesh, Paeiz Azmi,
   		 \textit{Senior Member, IEEE,}
     		 Nader Mokari, 
    		 \textit{Member, IEEE,}
     	 and Dennis Goeckel,  \textit{Fellow, IEEE} } 
     	  \textsuperscript{}\thanks{\noindent\textsuperscript{} Moslem Forouzesh is with the Department of Electrical and
  		Computer Engineering, Tarbiat Modares University, Tehran, Iran
  		(e-mail: m.Forouzesh@modares.ac.ir).
  		
  		Paeiz Azmi is with the Department of ECE, Tarbiat Modares University,
  		Tehran, Iran (e-mail: pazmi@modares.ac.ir).
  		
  		Nader Mokari is with the Department of ECE, Tarbiat Modares University,
  		Tehran, Iran (e-mail: nader.mokari@modares.ac.ir).
  
  D. Goeckel is with the Electrical and Computer Engineering Department, University of Massachusetts, Amherst, Massachusetts (e-mail:
  goeckel@ecs.umass.edu).	
  }}
 
 \maketitle
  \begin{abstract}
The study of the fundamental limits of covert communications, where a transmitter Alice wants to send information to a desired recipient Bob without detection of that transmission by an attentive and capable warden Willie, has emerged recently as a topic of great research interest. Critical to these analyses is a characterization of the detection problem that is presented to Willie. Previous work has assumed that the channel distribution information (CDI) is known to Alice, hence facilitating her characterization of Willie's capabilities to detect the signal. However, in practice, Willie tends to be passive and the environment heterogeneous, implying a lack of signaling interchange between the transmitter and Willie makes it difficult if not impossible for Alice to estimate the CDI exactly and provide covertness guarantees. In this paper, we address this issue by developing covert communication schemes for various assumptions on Alice's imperfect knowledge of the CDI: 1) when the transmitter knows the channel distribution is within some distance of a nominal channel distribution; 2) when only the mean and variance of the channel distribution are available at Alice; 3) when Alice knows the channel distribution is complex Gaussian but the variance is unknown. In each case, we formulate new optimization problems to find the power allocations that maximize covert rate subject to a covertness requirement under uncertain CDI. Moreover, since Willie faces similar challenges as Alice in estimating the CDI, we investigate two possible assumptions on the knowledge of the CDI at Willie: 1) CDI is known at Willie, 2) CDI is unknown at Willie. Numerical results are presented to compare the proposed schemes from various aspects, in particular the accuracy and efficiency of the proposed solutions for attaining desirable covert system performance.
 
 \end{abstract}

\section{Introduction}
Security and privacy are vital issues in wireless communications but are made challenging by the broadcast nature of the medium. To address these challenges, cryptographic and information-theoretic secrecy approaches have been considered, generally to protect a message's content from being obtained by eavesdroppers. However, in some scenarios, we want not only to protect the message content but also prevent an adversary from knowing that the message even exists \cite{AWGN_ch}, \cite{BSC}; that is, transmitter Alice wants to reliably send a message to intended recipient Bob without detection of the mere {\em presence} of that message by a capable and attentive adversary warden Willie. This method which hides the communication between nodes has been termed ``covert'' communication \cite{BSC}, \cite{DMC}.

Despite great historical interest in hiding a signal in low probability of detection (LPD) systems, the fundamental limits of covert communications were considered only relatively recently in \cite{AWGN_ch}. Since that time, the fundamental limits of covert communication have attracted significant research interest. In particular, covert communication has been studied for the additive white Gaussian noise (AWGN) channel, binary symmetric channel, and discrete memoryless channel in \cite{AWGN_ch}, \cite{BSC}, and \cite{DMC}, respectively. The authors in \cite{jammer} investigate covert communication in the presence of a friendly jammer, in which the jammer transmits a signal to aid covert communication by confusing the warden Willie about the nature of the background environment. Covert communication in one-way relay networks is studied in \cite{Hu.J}. A comparison between covert communications and information-theoretic security from different points of view is studied in \cite{Forouzesh}. In \cite{S.Y}, covert communications is considered for quasi-static wireless fading channels when artificial noise (AN) generated by a full-duplex receiver is employed.

The provisioning of covert communications relies critically on an accurate performance characterization of warden Willie's detector. In order to evaluate the detection error probability at Willie, the probability distribution function of the channel coefficient affecting the transmission from Alice to Willie should be known at Alice. However, in covert communications, Willie tends to be passive (it does not transmit a jamming signal), in part because its jamming signal can aid Alice's covertness \cite{jammer}. Hence, the study of covert communication under partial and uncertain channel distribution information (CDI) is important; if the transmitter makes the wrong assumption, covertness is lost. Note that prior studies on covert communication (e.g. \cite{jammer}) assume that the transmitter knows the CDI for the channel between Alice and Willie. In contrast, uncertain CDI has been investigated in different fields. For example, in \cite{ioannou2012outage}, the authors investigate the outage probability of a class of block fading Multi-Input Multi-Output (MIMO) channels. Effects of imperfect CDI on the performance of wireless networks is studied in \cite{N.Mokari}. Robust transmit power and bandwidth allocation for cognitive systems is evaluated in \cite{R.Fan} when partial information of an interference channel is known. To the best of our knowledge, there is no work on the important topic of uncertain CDI in covert communications.

In this paper, we investigate covert communication in the presence of the transmitter Alice, a jammer, a legitimate receiver Bob, and warden Willie, with all nodes equipped with a single antenna. In contrast to \cite{jammer}, the CDI for the channel between Alice and Willie is uncertain at Alice. Since different characterizations of the CDI may be available in different applications and systems, we consider the following cases:

\begin{enumerate}
	
	\item
	\underline{Nominal probability density function (N-pdf)}
	\begin{itemize}
		\item Availability of a nominal probability density function (pdf) of the channel. 
		\item We investigate a situation in which perfect CDI is not available at legitimate transmitters. In this scenario, only a class centered on the nominal distribution exists and the real CDI is a member of that class.
	\end{itemize}
	\item
	\underline{Mean and variance (MV)}
	
	\begin{itemize}
	\item  Availability of only the mean and variance of the channel distribution.
	\item 	In some situations, estimation of partial statistical information, such as the mean and variance, is practical and easier than the estimation of the channel distribution. In this case, we employ probabilistic inequalities.  
\end{itemize}
	
	\item
	\textcolor{black}{\underline{Full CDI with unknown variance (FCDI-UV) }}
	
	\begin{itemize}
	\item  Availability of the channel distribution with unknown variance. 
	\item  	We investigate a situation in which legitimate transmitters know Willie's channel distribution is complex Gaussian, but the variance is unknown. In this case, we evaluate the worst case scenario.
\end{itemize}
	
\end{enumerate}

Our aim is to maximize the covert rate subject to satisfying the covert communication requirements for the three proposed scenarios.  We investigate two possible assumptions for Willie: 1) Willie knows CDI perfectly, 2) Willie knows CDI imperfectly.  Our key contributions are summarized as follows:

\begin{itemize}
	
	\item
	
	We present novel optimization problems for solving the covert communications problem when Alice has uncertainty about the CDI on the channel from herself to warden Willie.
		
	\item
	
	We devise new iterative algorithms to solve the proposed optimization problems.

	\item
	
	We provide extensive numerical results to characterize the performance of the proposed algorithms.
	
\end{itemize}

 The remainder of this paper is organized as follows: In Section \ref{system model}, the system model and covertness metrics are presented. Section \ref{Optimization Problem} provides the problem formulations. Power optimization for the three different cases of uncertainty for the CDI at Willie and the legitimate parties are investigated in Sections \ref{CUW}. Moreover, assumptions of certain CDI at Willie and uncertain CDI at legitimate parties are studied in Section \ref{CKW}. Numerical results and performance evaluation are presented in Section \ref{Numerical Results}. Finally, the conclusions of the paper are presented in Section \ref{Conclusion}.

 \section{System Model}\label{system model}
\begin{figure}[h!]
  		\includegraphics[width=3in,height=3in]{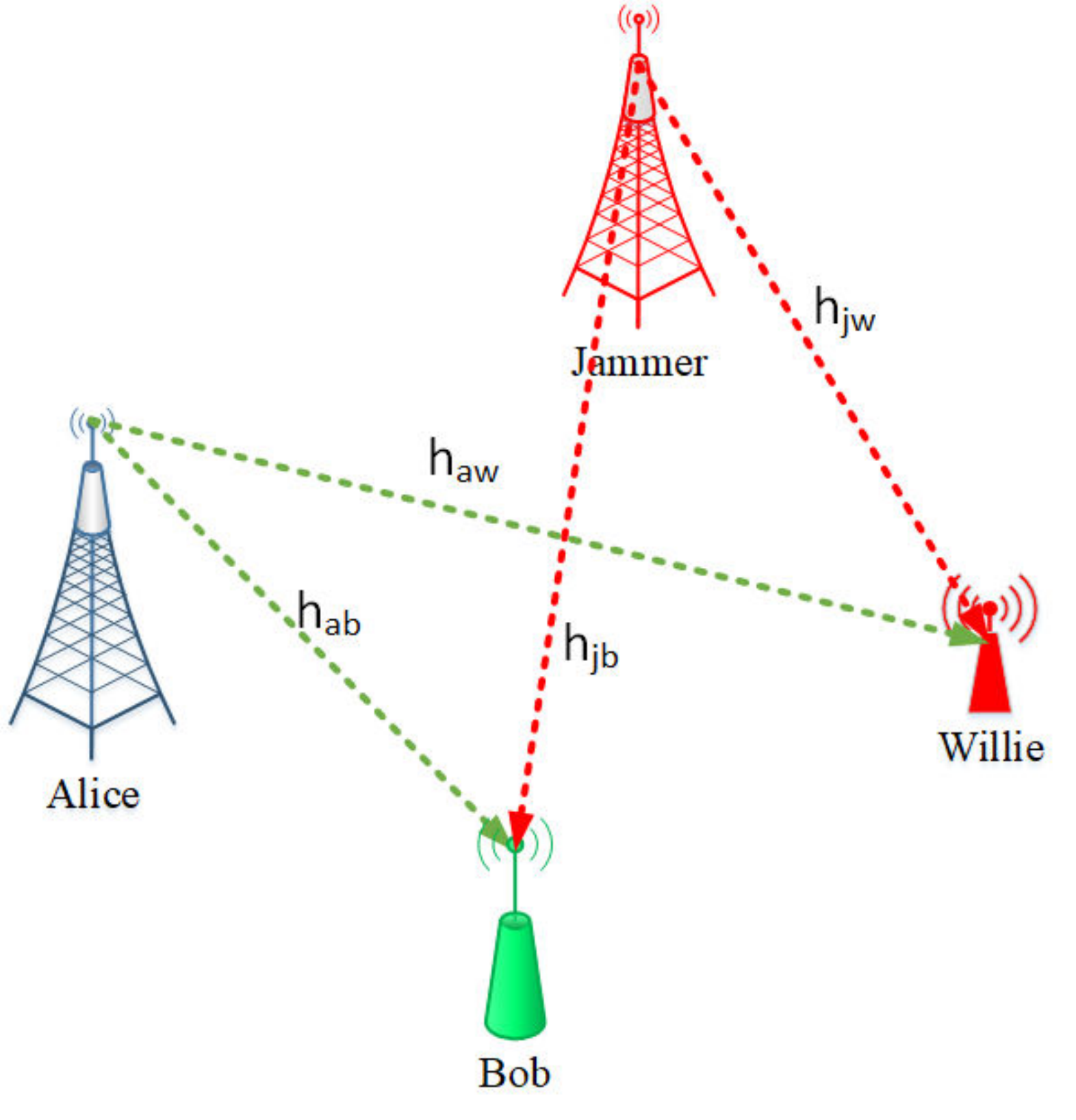}
 		\caption{System model:  Legitimate transmitter Alice attempts to communicate with legitimate receiver Bob without the detection of that message by capable and attentive warden Willie in the presence of a jammer.}
 		\label{Sys}
  \end{figure}

 We consider  a system model which consists of Alice, Bob, a jammer, and a single Willie, as shown in
 Fig. \ref{Sys}. Alice, Bob, the jammer and Willie are each equipped with a single antenna.  Alice attempts to transmit private
 messages to Bob covertly, and the jammer broadcasts the jamming signal to confuse Willie. This communication occurs in a discrete-time channel with $\Omega $ time slots, and the length of each time slot is $n$ symbols. The private message and jamming signal  in each time slot can be written as  ${{\boldsymbol{x}}^d} = \left[ {x^d_1,x^d_2,\dots,x^d_n} \right]$ and  ${{\boldsymbol{x}}^j} = \left[ {x^j_1,x^j_2,\dots,x^j_n} \right]$, respectively. 
We define the distance from Alice to Bob, Alice to Willie, the jammer to Bob, and the jammer to Willie as $d_{ab}$, $d_{aw}$, $d_{jb}$, and $d_{jw}$, respectively. Moreover, $h_{ab}$, $h_{aw}$, $h_{jb}$, and $h_{jw}$ are the channel coefficients for the channel from Alice to Bob, Alice to Willie,  the jammer to Bob, and jammer to Willie, respectively.

The problem can be easily solved if Alice transmits both data and jamming, or Alice informs the jammer about the timing of her transmission \cite{jammer}. In particular, if Alice and the jammer each draw their transmissions from the same distribution and the jammer uses knowledge of the timing of Alice's transmission to silence the jammer transmission whenever Alice transmits, the statistics of the received signal at Willie are the same whether Alice is transmitting or not, since $\boldsymbol{x}^d$ and $\boldsymbol{x}^j$ will have identical statistics.  Accordingly, Willie cannot detect the covert transmission and hence the covert communication problem is readily solved.   Hence, in this paper (as in \cite{jammer}), we investigate the more interesting case in which Alice and the jammer do not interact.

Because we want to make certain that covertness is not compromised, we pessimistically assume that legitimate transmitters have imperfect CDI for the channels between system nodes and Willie.  In particular, although the legitimate nodes might be communicating with each other in a similar propagation environment, the environment might be heterogeneous and thus the lack of a signaling interchange between the legitimate transmitters and Willie make it difficult for the legitimate transmitter to estimate this CDI exactly. Moreover, because the warden Willie can be faced with the same challenge due to the covertness of Alice's transmission, it is necessary to study two possible assumptions: 1) Willie knows the CDI of  the Alice-to-Willie and the jammer-to-Willie channels perfectly, 2) Willie knows the CDI of  Alice-to-Willie and the jammer-to-Willie channels imperfectly.
 
 The received signal at  receiver $m$ (Bob or Willie) is given by \cite{jammer}:
\begin{equation}\label{Received signal}
{\bf{y}}_m = \left\{ {\begin{array}{*{20}{l}}
	{\frac{{\sqrt {{p_j}} {h_{jm}}{{\bf{x}}^j}}}{{d_{jm}^{\alpha /2}}} + {{\bf{z}}_m},}&{{\Psi _0},}\\
	{\frac{{\sqrt {{p_j}} {h_{jm}}{{\bf{x}}^j}}}{{d_{jm}^{\alpha /2}}} + \frac{{\sqrt {{p_d}} {h_{am}}{{\bf{x}}^d}}}{{d_{am}^{\alpha /2}}} + {{\bf{z}}_m},}&{{\Psi _1},}
	\end{array}} \right.
\end{equation}
where $p_{d}$ and $p_{j}$ are the transmit power for the data and the jamming signals, respectively, where ${p_{d}} = \rho {P}$, $
{p_j} = \left( {1 - \rho } \right){P}$ with $\rho$ defined as the power allocation factor, $\alpha$  is the path-loss exponent, and $\boldsymbol{z}_m \sim \mathcal{CN}\left( {\bf{0},\sigma _m^2} {\bf I}_n\right)$ represents the complex Gaussian received noise vector at user $m$, where ${\bf{I}}_n$
is the $n \times n$  identity matrix and $\bf{0}$  is the length-$n$ zero vector. Moreover, $P$ indicates the total transmit power and $\boldsymbol{y}_m=\left[ {y_m^1,y_m^2, \ldots y_m^\iota ,...,y_m^n} \right]$.  The events $\Psi_0$ and $\Psi_1$ indicate that Alice does not transmit data to Bob and transmits data to Bob, respectively.

Without loss of  generality, we assume that Alice and the jammer employ Gaussian codebooks and Gaussian jamming, respectively \cite{jammer}. Hence, According to \eqref{Received signal}, the conditional distribution of each element of the received vector at Willie given $h_{aw}$ and $h_{jw}$ has a complex Gaussian distribution with zero mean and variance $\sigma _w^2 + \gamma$, i.e.,
$ y_w^\iota | h_{aw},h_{jw}  \sim  {\cal C}{\cal N}\left( {0,\sigma _w^2 + \gamma } \right)
$, where $\gamma$ is defined as:
\begin{align}\label{gamma}
\gamma  = \left\{ {\begin{array}{*{20}{l}}
	{\frac{{\left( {1 - \rho } \right){P}{{\left| {{h_{jw}}} \right|}^2}}}{{d_{jw}^\alpha }},}&{{\Psi _0},}\\
	{\frac{{\left( {1 - \rho } \right){P}{{\left| {{h_{jw}}} \right|}^2}}}{{d_{jw}^\alpha }} + \frac{{\rho {P}{{\left| {{h_{aw}}} \right|}^2}}}{{d_{aw}^\alpha }},}&{{\Psi _1}.}
	\end{array}} \right.
\end{align}
When Willie decides $\Psi_1$ while $\Psi_0$ is true, a False Alarm (FA) with probability $\mathbb{P}_{FA}$ has occurred. If  Willie decides $\Psi_0$ while Alice transmits data to Bob, the Miss Detection (MD) with probability $\mathbb{P}_{MD}$ has occurred.
As in prior work in covert communications, we adopt the covertness criterion of \cite{AWGN_ch}.  Alice has the covert communication to Bob when the following constraint is satisfied for $\varepsilon > 0$ \cite{jammer}:
\begin{align}\label{decision}
\mathbb{P}_{MD}+\mathbb{P}_{FA}\ge 1-\varepsilon, \,\,\, \text{as}\,\,\, n \to \infty.
\end{align}
We assume a capable Willie, which means that he employs a well-considered decision rule to minimize his detection error.  In the case that Willie knows the fading coefficient $h_{a,w}$ of the Alice-to-Willie channel and the CDI of the jammer-to-Willie channel perfectly, the optimal decision rule at Willie for our scenario can be shown to be a power detector \cite{jammer}:
\begin{align}\label{Radi_mt}
\frac{{{\Re}}}{n}	\mathop \gtrless\limits_{\Psi_0}^{\Psi_1} \vartheta,
\end{align}
where $\vartheta$ is the decision threshold and $\Re$ denotes the total received power at Willie in each time slot; that is, $\Re={\sum\limits_{\iota   = 1}^n {\left| {y_w^\iota  } \right|} ^2}$.
  In the case where Willie does not know $h_{aw}$ or the CDI of the Alice-to-Willie channel perfectly, the result of \cite{jammer} no longer applies and it is difficult to establish the optimal decision rule at Willie; however, a power detector is still likely to be employed and is often adopted in the literature, \cite{S.Y}, \cite{S.Yan}.
Therefore, we can formulate the  FA and MD probabilities as follows:
\begin{align}\label{Ps}
&{\mathbb{P}_{FA}} = \nonumber\\& \mathbb{P}\left( {\frac{{{\sum\limits_{\iota   = 1}^n {\left| {y_w^\iota  } \right|} ^2}}}{n} > \vartheta \left| {{\Psi_0}} \right.} \right),\,\,\,
{\mathbb{P}_{MD}} = \mathbb{P}\left( {\frac{{{\sum\limits_{\iota   = 1}^n {\left| {y_w^\iota  } \right|} ^2}}}{n} < \vartheta \left| {{\Psi_1}} \right.} \right).
\end{align}
As $
y_w^\iota  | h_{aw},h_{jw}  \sim   {\cal C}{\cal N}\left( {0,\sigma _w^2 + \gamma } \right)
$, we have  ${\sum\limits_{\iota   = 1}^n {\left| {y_w^\iota  } \right|} ^2}| h_{aw},h_{jw}  \sim \left( {\sigma _w^2 + \gamma } \right) \chi _{2n}^2$, where $\chi _{2n}^2$ is a chi-squared random variable with $2n$ degrees of freedom. 
Hence, 
 \eqref{Ps} can be rewritten as follows:
\begin{align}\label{PFA1}
&{\mathbb{P}_{FA}} = \nonumber\\& \mathbb{P}\left( {\frac{{{\sum\limits_{\iota   = 1}^n {\left| {y_w^\iota  } \right|} ^2}}}{n} > \vartheta \left| {{\Psi_0}} \right.} \right) = \mathbb{P}\left( {\left( {\sigma _w^2 + \gamma } \right)\frac{{\chi _{2n}^2}}{n} > \vartheta \left| {{\Psi_0}} \right.} \right),
\end{align}
and
\begin{align}\label{PMD1}
&{\mathbb{P}_{MD}} = \nonumber\\&\mathbb{P}\left( {\frac{{{\sum\limits_{\iota   = 1}^n {\left| {y_w^\iota  } \right|} ^2}}}{n} < \vartheta \left| {{\Psi_1}} \right.} \right) = \mathbb{P}\left( {\left( {\sigma _w^2 + \gamma } \right)\frac{{\chi _{2n}^2}}{n} < \vartheta \left| {{\Psi_1}} \right.} \right). 
\end{align}
Much of the literature in covert communications has considered performance as a function of $n$, which captures the effect of the rate of convergence of power measurements at Willie.  However, it is also instructive to take an ``outage'' approach, as has been considered in \cite{S.Y}, \cite{S.Ya}.  In the outage approach, the dependence on $n$ is suppressed by letting $n \rightarrow \infty$, and we consider the probability that the channel conditions are such that covertness or reliability is obtained when Willie has a perfect estimate of the power at his receiver.  This often captures the salient aspects of the problem while more clearly illustrating the underlying mechanisms; for example, the achievability results in \cite{jammer} can be readily obtained in a more transparent fashion by first taking $n \rightarrow \infty$ to give Willie a perfect estimate of his received power and then performing the analysis.

To take this outage approach, we let $n \rightarrow \infty$ and employ the Strong Law of Large Numbers (SLLN) to obtain $\frac{{\chi _{2n}^2}}{n} \rightarrow 1$.  Based on Lebesgue's Dominated Convergence Theorem \cite{coverage}, $\frac{\chi _{2n}^2}{n}$  can be replaced with 1, as $n \to \infty $. Hence, we can rewrite \eqref{PFA1} and \eqref{PMD1} as follows: 
\begin{align}\label{pfa}
{\mathbb{P}_{FA}}& = \mathbb{P}\left( {\left( {\sigma _w^2 + \gamma } \right)\frac{{\chi _{2n}^2}}{n} > \vartheta \left| {{\Psi_0}} \right.} \right)= \mathbb{P}\left( {\sigma _w^2 + \gamma  > \vartheta \left| {{\Psi _0}} \right.} \right), \nonumber\\& = \mathbb{P} \left( {\sigma _w^2 + \frac{{{{\left| {{h_{jw}}} \right|}^2}}}{{d_{jw}^\alpha }}\left( {1 - \rho } \right){P} > \vartheta } \right),{\rm{ }} \nonumber\\& = \mathbb{P} \left( {{{\left| {{h_{jw}}} \right|}^2} > \frac{{\left( {\vartheta  - \sigma _w^2} \right)d_{jw}^\alpha }}{{\left( {1 - \rho } \right){P}}}} \right),
\end{align}
\begin{align}\label{pmd}
&{\mathbb{P}_{MD}}=\mathbb{P}\left( {\left( {\sigma _w^2 + \gamma } \right)\frac{{\chi _{2n}^2}}{n} < \vartheta \left| {{\Psi_1}} \right.} \right) = \mathbb{P}\left( {\sigma _w^2 + \gamma  < \vartheta \left| {{\Psi _1}} \right.} \right), \nonumber\\&  =\mathbb{P} \left( {\sigma _w^2 + \frac{{{{\left| {{h_{aw}}} \right|}^2}}}{{d_{aw}^\alpha }}\rho {P} + \frac{{{{\left| {{h_{jw}}} \right|}^2}}}{{d_{jw}^\alpha }}\left( {1 - \rho } \right){P} < \vartheta } \right).{\rm{ }}
\end{align}

\section{ Optimization Problem}\label{Optimization Problem}
We aim to maximize the covert rate when there is uncertainty about the CDI between legitimate transmitters and Willie, subject to a transmit power limitation and a covert communication constraint in \eqref{decision}. Hence,  the following optimization problem is considered:
\begin{subequations}\label{Opt}
	\begin{align}
	&\max_{\rho}\;\mathbb{P}_{\psi_1}  \log \left( {1 + \frac{{\rho {\gamma _b}}}{{1 + \left( {1 - \rho } \right){\gamma _j}}}} \right),\\& \label{Opti_prob1}
	\hspace{.15cm} \text{s.t.}:\hspace{.18cm} 0\le \rho\le1,\\&\label{Opti_prob2}\hspace{.85cm} 
	\hspace{.008cm}\min_{\vartheta}\left(\;{\mathbb{P}_{FA}} + {\mathbb{P}_{MD}}\right)\ge 1-\varepsilon,
	\end{align}
\end{subequations}
where $\mathbb{P}_{\psi_1}$, ${\gamma _b}$, and ${\gamma _j}$ represent the probability of transmission of data  by Alice in each time slot, ${\gamma _b} = \frac{{{P}{{\left| {{h_{ab}}} \right|}^2}}}{{d_{ab}^\alpha \sigma _b^2}}$, and ${\gamma _j} = \frac{{{P}{{\left| {{h_{jb}}} \right|}^2}}}{{d_{jb}^\alpha \sigma _b^2}}$, respectively.  Constraint  \eqref{Opti_prob1} represents upper and lower bounds on the power allocation factor, $\rho$. Constraint \eqref{Opti_prob2} represents  the requirement of covert communication based on the worst case threshold from the point of view of Alice. َ
As mentioned before, for investigation of imperfect CDI, we study three scenarios with different assumptions: 1) N-pdf: Availability of a nominal probability density function of the channel. In this scenario, we investigate a situation in which  perfect CDI is not available at the legitimate transmitters. Rather a  class  centered around a nominal distribution exists, and the desired channel distribution is a member of that class. This scenario might arise when the CDI is calculated based on  multiple but limited observations. Moreover, when legitimate transmitters cannot estimate the channel distribution in the current time slot, this scenario is useful, with the nominal distribution being the channel distribution estimated in the prior time slot. 
 2) MV:‌	Availability of only mean and variance of channel distribution. In some situations, estimation of some statistical information such as the mean and variance are practical and easier than estimation of the full channel distribution. In this scenario, we assume legitimate transmitters only know mean and variance of channel distribution.
  3) FCDI-UV:	Availability of the  distribution with unknown variance. In this scenario, we investigate a situation in which legitimate transmitters know the channel distribution but the variance estimation
  is unknown. 
    In the following, we study these schemes for two assumptions at Willie: 1) CDI is known at Willie, 2) CDI is unknown at Willie.

\section{Scenario I: CDI is Unknown at Willie }\label{CUW}
In this scenario, we assume that the environment is heterogeneous or changing rapidly; hence, 
 Willie does not know the CDI perfectly due to similar problems as the legitimate transmitters in obtaining such.  In this case, since the CDI is uncertain at Willie, he should calculate the decision threshold that minimizes the worst case detection error from his point of view. And hence Alice will allocate transmit power under this strategy at Willie.   Thus, \eqref{Opti_prob2} is replaced with $\min\limits_{\vartheta}\mathop {\min }\limits_{\begin{array}{*{20}{l}}
	_{f_w} , _{f_j}
	\end{array}}\left(\;{\mathbb{P}_{FA}} + {\mathbb{P}_{MD}}\right)\ge 1-\varepsilon.$ 
	\textcolor{black}{
 \subsection{N-pdf Scenario} \label{M1}
 Legitimate transmitters do not know the perfect CDI and only know that a  class  centered around a nominal distribution exists, and the channel distributions are members of it.  
In particular, in this scenario, legitimate transmitters know that the channel coefficient probability density functions (pdf)  $f_{{{{\left| {{h_{aw}}} \right|}^2}}}\left( {x} \right)$ and $f_{{{\left| {{h_{jw}}} \right|}^2}}\left( {y} \right)$ are within a specified distance of the nominal distributions $f_{_{0,{{\left| {{h_{aw}}} \right|}^2}}}\left( {x} \right)$ and $f_{_{0,{{\left| {{h_{jw}}} \right|}^2}}}\left( {y} \right)$, respectively. We employ relative entropy to measure the distance between two distributions. Therefore,  the following inequality  is satisfied by a feasible distribution $f_{_{{{\left| {{h_{aw}}} \right|}^2}}}\left( {x} \right)$  \cite{Thomas}: 
\begin{align}\label{RA}
&D\left( {f_{{{\left| {{h_{aw}}} \right|}^2}}\left( {x} \right)\left\| {{f_{_{0,{{\left| {{h_{aw}}} \right|}^2}}}}(x)} \right.} \right) =\\& \int {f_{{{\left| {{h_{aw}}} \right|}^2}}(x)\ln \frac{f_{{{\left| {{h_{aw}}} \right|}^2}}(x)}{{{f_{_{0,{{\left| {{h_{aw}}} \right|}^2}}}(x)}}}dx}  \le \ell \nonumber
\end{align} 
 where $D\left( {f_{{\left| {{h_{aw}}} \right|}^2}(x)\left\| {f_{_{0,{\left| {{h_{aw}}} \right|}^2}}\left( {x} \right)} \right.} \right) $ is defined as the relative entropy between two pdfs $f_{{{\left| {{h_{aw}}} \right|}^2}}(x) $ and  $f_{_{0,{\left| {{h_{aw}}} \right|}^2}}(x) $, and $\ell$ is the maximum distance from the nominal distribution. In this scenario, legitimate transmitters know  $\ell$ and ${f_{_{0,{\left| {{h_{aw}}} \right|}^2}}(x)}$.   Likewise, we have the analog of \eqref{RA}  for  $f_{{\left| {{h_{jw}}} \right|}^2}\left( {y} \right)$ and $f_{_{0,{\left| {{h_{jw}}} \right|}^2}}\left( {y} \right)$. In this paper, for simplification of the notation, we use $f_a$, $f_j$, $f_{_{0,a}}$, $f_{_{0,j}}$,  $h_a$, and $h_j$ instead of $f_{{\left| {{h_{aw}}} \right|}^2}\left( {x} \right)$, $f_{{\left| {{h_{jw}}} \right|}^2}\left( {y} \right)$,  $f_{_{0,{\left| {{h_{aw}}} \right|}^2}}\left( {x} \right)$, $f_{_{0,{\left| {{h_{aw}}} \right|}^2}}\left( {y} \right)$, ${{{\left| {{h_{aw}}} \right|}^2}}$, and ${{{\left| {{h_{jw}}} \right|}^2}}$, respectively.}

	\textcolor{black}{
In this scenario, our optimization problem is formulated as follows: 
 \begin{subequations}\label{Opt_method1}
	\begin{align}
&\max_{\rho}\;\mathbb{P}_{\psi_1}  \log \left( {1 + \frac{{\rho {\gamma _b}}}{{1 + \left( {1 - \rho } \right){\gamma _j}}}} \right),\label{Opt_method1_0}\\& 
\hspace{.15cm}\text{s.t.}: \hspace{.18cm}  \eqref{Opti_prob1}, \nonumber
\\&\hspace{.85cm}  \min_{\vartheta}\mathop {\min }\limits_{\begin{array}{*{20}{l}}
	{_{D\left( {{f_a}\left\| {{f_{0,a}}} \right.} \right) \le \ell}}\\
	{_{D\left( {{f_j}\left\| {{f_{0,j}}} \right.} \right) \le \ell }}
	\end{array}}\left(\;{\mathbb{P}_{FA}} + {\mathbb{P}_{MD}}\right)\ge 1-\varepsilon,\label{Opt_method1_2}
\end{align}
\end{subequations}
where \eqref{Opt_method1_2} demonstrates the worst case covert communication requirement from the point of view of the legitimate transmitters.
 For solving \eqref{Opt_method1},
 first we investigate \eqref{Opt_method1_2}. The inner minimum in the left side of \eqref{Opt_method1_2} cannot be solved directly. Hence, we again are pessimistic from Alice's perspective and consider a lower bound for this minimization.  Define random variable $W = \frac{{\rho {P}{{\left| {{h_{aw}}} \right|}^2}}}{{d_{aw}^\alpha }} + \frac{{\left( {1 - \rho } \right){P}{{\left| {{h_{jw}}} \right|}^2}}}{{d_{jw}^\alpha }}$, which leads to ${\mathbb{P}_{MD}}=\mathbb{P} \left( {\sigma _w^2 + W < \vartheta } \right).{\rm{ }}$  Finally,  the lower bound of the inner minimum \eqref{Opt_method1_2} can be written as: $\mathop {\min }\limits_{D\left( {{f_j}\left\| {{f_{0,j}}} \right.} \right) \le \ell }\mathbb{P}_{FA}  + \mathop {\min }\limits_{D\left( {{f_w}\left\| {{f_{0,w}}} \right.} \right) \le \ell } \mathbb{P}_{MD} $, where $f_{_W}$ and $f_{_{0,W}}$  are the simplified notations of $f_{_W}(w)$ and $f_{_{0,W}}(w)$, respectively, $f_{_W}$ is the desired pdf of random variable $W$ and
  $f_{_{0,W}}(w)={\frac{{d_{aw}^\alpha }}{{\rho P}}}f_{_{0,{\left| {{h_{aw}}} \right|}^2}}\left( {{\frac{{d_{aw}^\alpha {w}}}{{\rho P}}}} \right)*\frac{{d_{jw}^\alpha }}{{\left( {1 - \rho } \right)P}}
  f_{_{0,{\left| {{h_{jw}}} \right|}^2}}\left( { {\frac{{d_{jw}^\alpha {w}}}{{ {(1 - \rho) } P}}} } \right)$,  where $*$ is the convolution operator.
   The reason is that for given $g_1$ and  $g_3$, there is unique $g_2$ such that $g_1=g_2*g_3$. Finally, in order to find the optimal $f_j$ and $f_w$, we should solve the following optimization problems:
 \vspace{-.5cm}
\begin{multicols}{2}
 \begin{subequations}\label{covert_req}
	\begin{align}
	&\min_{f_j}\;\;{\mathbb{P}_{FA}},\label{covert_req0}\\& 
	\hspace{-.1cm}\text{s.t.}: \hspace{.05cm} \int\limits_{}^{} {f_j\ln \left( {\frac{f_j}{{f_{_{0,j}}}}} \right)dy}  \le \ell, \label{covert_req1}
	\\&\hspace{.36cm}  \int\limits_{}^{} f_j dy = 1. \label{covert_req2}	
	\end{align}
\end{subequations}
 \begin{subequations}\label{covert_req_1}
	\begin{align}
	&\min_{f_w}\;\; {\mathbb{P}_{MD}},\label{covert_req_10}\\& 
	\hspace{-.1cm}\text{s.t.}: \hspace{.05cm} \int\limits_{}^{} {f_w\ln \left( {\frac{f_w}{{f_{_{0,w}}}}} \right)dw}  \le \ell, \label{covert_req_11}
	\\&\hspace{.36cm}  \int\limits_{}^{} f_w dw = 1. \label{covert_req_12}	
	\end{align}
\end{subequations}
\end{multicols}
By the definition  of ${\xi _0}{\left( y \right)}$ and ${\xi _1}{\left( w \right)}$ as indicators of the FA and MD complementary sets, i.e., 
${\xi _0}\left( {{y}} \right) = \left\{ {\begin{array}{*{20}{l}}
	1,&{y < \frac{{\left( {\vartheta  - \sigma _w^2} \right)d_{jw}^\alpha }}{{\left( {1 - \rho } \right){P}}}},\\
	0,&{\mbox{else}},
	\end{array}} \right.$
 and
  ${\xi _1}\left( w \right) = \left\{ {\begin{array}{*{20}{l}}
  	1,&{w > \left( {\vartheta  - \sigma _w^2} \right)},\\
  	0,&{\mbox{else}},
  	\end{array}} \right.$
   respectively, we can reformulate \eqref{covert_req} and \eqref{covert_req_1} as follows
    \vspace{-.9cm}}
   \begin{multicols}{2}
   		\textcolor{black}{	\begin{align}\label{CRj}
   		&\min_{f_j}\;\;{1-\int\limits_{}^{} {{\xi _0}{\left(y \right)}} f_j dy},\nonumber\\& 
   		\hspace{-.1cm}\text{s.t.}: \hspace{.05cm} \eqref{covert_req1},  \eqref{covert_req2}, 	
   		\end{align}}
   		
   			\textcolor{black}{
   		\begin{align}\label{CRw}
   		&\min_{f_w}\;\; 1-\int\limits_{}^{} {{\xi _1}{\left( w \right)}} f_w dw, \nonumber\\& 
   		\hspace{-.1cm}\text{s.t.}: \hspace{.05cm}    \eqref{covert_req_11},  \eqref{covert_req_12}.	
   		\end{align}}
   \end{multicols}
	\textcolor{black}{In order to obtain the optimal $f_j$ and $f_w$, it is sufficient to solve the following equivalent optimization problems
   \begin{multicols}{2}
	\begin{align}\label{CR_j}
	&\max_{f_j}\;\;{\int\limits_{}^{} {{\xi _0}{\left(y \right)}} f_j dy},\nonumber\\& 
	\hspace{-.1cm}\text{s.t.}: \hspace{.05cm} \eqref{covert_req1},  \eqref{covert_req2}, 	
	\end{align}
\vspace{-.7cm}
	\begin{align}\label{CR_w}
	&\max_{f_w}\;\; \int\limits_{}^{} {{\xi _1}{\left( w \right)}} f_w dw, \nonumber\\& 
	\hspace{-.1cm}\text{s.t.}: \hspace{.05cm}    \eqref{covert_req_11},  \eqref{covert_req_12}.	
	\end{align}
\end{multicols}
The solution to the above optimization problems has already been proposed in  \cite{ioannou2012outage};  consequently, we can write the worst-case channel distributions $f_j^*$ and  $f_w^*$ as follows, \cite{ioannou2012outage}:
	\begin{align}
f_j^* = \frac{{\left( {{e^{\frac{1}{{\beta _j^*}}}} - 1} \right){\xi _0}\left( {{y}} \right) + 1}}{{\left( {{e^{\frac{1}{{\beta _j^*}}}} - 1} \right){\Im _j} + 1}}{f_{_{0,j}}},
\end{align}
	\begin{align}
f_w^* = \frac{{\left( {{e^{\frac{1}{{\beta _w^*}}}} - 1} \right){\xi _1}\left( {{w}} \right) + 1}}{{\left( {{e^{\frac{1}{{\beta _w^*}}}} - 1} \right){\Im _w} + 1}}{f_{_{0,w}}},
\end{align}
where ${\Im _j} = \int {{\xi _0}\left( {{y}} \right){f_{_{0,j}}}} d{y}$ and ${\Im _w} = \int {{\xi _1}\left( w \right){f_{_{0,w}}}} dw$. Moreover,   $\beta_j^*$ and $\beta_w^*$ are given by \eqref{Bj} and \eqref{Ba}, respectively, at the top of the next page.}
 \begin{figure*}[t]    	 
 	\begin{align}
&\beta_j^*=\begin{cases}{ - \ln \left( {1 + \Im_j \left( {{e^{\frac{1}{\beta_j }}} - 1} \right)} \right) + \frac{{{e^{\frac{1}{\beta_j }}}\Im_j }}{{1 + \Im_j \left( {{e^{\frac{1}{\beta_j }}} - 1} \right)}}\frac{1}{\beta_j } = \ell, } & {\ell  < \ln \left( {1/\Im_j } \right)}\\0, &  \mbox{else} \end{cases} \label{Bj}\\&
 \beta_w^*=\begin{cases}{ - \ln \left( {1 + \Im_w \left( {{e^{\frac{1}{\beta_w }}} - 1} \right)} \right) + \frac{{{e^{\frac{1}{\beta_w }}}\Im_w }}{{1 + \Im_w \left( {{e^{\frac{1}{\beta_w }}} - 1} \right)}}\frac{1}{\beta_w } = \ell, } & {\ell  < \ln \left( {1/\Im_w } \right)}\\0, &  \mbox{else} \end{cases} \label{Ba} 
 	\end{align}
 	\hrule
 \end{figure*}
  
  In order to facilitate further analysis, we continue by assuming the nominal distributions for $h_{aw}$ and $h_{jw}$ are complex Gaussian with zero mean and unit variance, i.e., $h_{aw} \sim {\cal C}{\cal N}\left( {0,1} \right)$ and $h_{jw} \sim {\cal C}{\cal N}\left( {0,1} \right)$; hence,  the nominal distributions of ${\left| {{h_{aw}}} \right|}^2$ and ${\left| {{h_{jw}}} \right|}^2$ are exponential with parameter one, and we have:
 \begin{align}\label{inner_minimization_fa}
&\min_{D\left( {{f_j}\left\| {{f_{0,j}}} \right.}  \right) \le \ell}\;{\mathbb{P}_{FA}}={1-\int\limits_{}^{} {{\xi _0}{\left( y \right)}} f_j^{*} dy}= 1 - \int_0^{\frac{{\left( {\vartheta  - \sigma _w^2} \right)d_{jw}^\alpha }}{{\left( {1 - \rho } \right){P}}}} {f_j^*d{y}}\nonumber \\& = 1 - \frac{{{e^{\frac{1}{{\beta _j^*}}}}}}{{\left( {{e^{\frac{1}{{\beta _j^*}}}} - 1} \right){\Im _j} + 1}}\left( {1 - {e^{ - \frac{{\left( {\vartheta  - \sigma _w^2} \right)d_{jw}^\alpha }}{{\left( {1 - \rho } \right){P}}}}}} \right), 
 \end{align}
 and
  \begin{align}\label{inner_minimization_md}
 &\min_{D\left( {{f_w}\left\| {{f_{0,w}}} \right.}  \right) \le \ell}\;{\mathbb{P}_{MD}}={1-\int\limits_{}^{} {{\xi _1}{\left( w\right)}} f_w^{*} dw}= 1 - \int_{\vartheta  - \sigma _w^2}^{\infty} {f_w^*d{w}}= \nonumber\\&  1 + \frac{{{e^{\frac{1}{{\beta _w^*}}}}}}{{\left( {{e^{\frac{1}{{\beta _w^*}}}} - 1} \right){\Im _w} + 1}} \times \frac{{  1}}{{\rho d_{aw}^{ - \alpha } - \left( {1 - \rho } \right)d_{jw}^{ - \alpha }}}\times\nonumber\\&\left[ { - \rho d_{aw}^{ - \alpha }{e^{ - \frac{{\left( {\vartheta  - \sigma _w^2} \right)}}{{\rho {P}d_{aw}^{ - \alpha }}}}} + \left( {1 - \rho } \right)d_{jw}^{ - \alpha }{e^{ - \frac{{\left( {\vartheta  - \sigma _w^2} \right)}}{{\left( {1 - \rho } \right){P}d_{jw}^{ - \alpha }}}}}} \right].
 \end{align}
Next, we solve the outer minimization \eqref{Opt_method1_2}, which is reformulated as:
 \begin{align}\label{min_nu}
\mathop {\min }\limits_\vartheta  \left ( 2-\int\limits_{}^{} {{\xi _0}{\left( y \right)}} f_j^{*} dy -\int\limits_{}^{} {{\xi _1}{\left( w \right)}} f_w^{*} dw \right).
\end{align}
It is clear that the objective function \eqref{min_nu} with respect to $\vartheta$ is a quasiconvex function because of the convexity of its domain and all sublevel sets. Hence, the
 problem \eqref{min_nu} is a quasiconvex optimization problem. It can be
 solved via quasiconvex programming that can be performed by the bisection method \cite{Boyd}.
 Because of the convex representation  of the sublevel sets in the quasiconvex objective function in \eqref{min_nu}, we can reformulate this problem as follows
 \begin{align}\label{fin_nu}
& \text {Find} \,\,\,\vartheta \nonumber
 \\& 
 \hspace{-.2cm}\text{s.t.}: \hspace{.05cm}  E=2-\int\limits_{}^{} {{\xi _0}{\left( y \right)}} f_j^{*} dy -\int\limits_{}^{} {{\xi _1}{\left( w \right)}} f_w^{*} dw - \lambda \le 0,
 \end{align}
 where $\lambda \in \mathbb{R}$ and $\mathbb{R}$ is the set of real numbers.  We employ the bisection method as shown in Algorithm \ref{Alg_bs}.

 \begin{algorithm}[t] 
 	\caption{BISECTION METHOD FOR FINDING THE OPTIMAL $\vartheta$ in \eqref{min_nu}} \label{Alg_bs}
 	\begin{algorithmic}[1]
 		\STATE  \nonumber
 		Initialization:\\ \hspace{.2cm} Set $\eta \to 0 $, $l \le E \left(\vartheta=\vartheta^* \right)$, $u \ge E \left(\vartheta=\vartheta^* \right)$, where $l=0$, $u=2$,
 		\STATE \label{find}
 		 $\lambda=\frac{l+u}{2}$,
 		\STATE  		
 		Solve the convex feasibility problem \eqref{fin_nu},
 		\STATE 
 		If \eqref{fin_nu} is feasible,
 		\\\hspace{.2cm} $u=\lambda$,\\
 		else
 		\\\hspace{.2cm} $l=\lambda$,
 		\STATE 
 	      If $\left| {u-l} \right| \le \eta $,\\
 		\hspace{.2cm} Stop,\\
 		else\\
 		\hspace{.2cm} Go back to step \ref{set}.
 	\end{algorithmic}
 \end{algorithm}	
 
 Since the logarithmic function is increasing,  we can maximize $g(x)$ instead of $\log(1 + g(x))$. Therefore, we maximize $g(\rho)=\frac{{\rho {\gamma _b}}}{{1 + \left( {1 - \rho } \right){\gamma _b}}} $ instead of the objective function in \eqref{Opt_method1}. Moreover, $g(\rho)$ is equivalent to $z(\rho)= \frac{1}{{\frac{1}{{\rho {\gamma _b}}} + \frac{1}{\rho } - 1} }$. Finally, we can minimize the denominator of $z(\rho)$ instead of maximizing $g(\rho)$. Therefore, the optimization problem can be reformulated as follows:
 	\begin{subequations}\label{MethodI_rho}
 	\begin{align}
 	&\min_{\rho}\; {\rho ^{ - 1}} + {\rho ^{ - 1}}{\gamma _b},\label{}\\& 
 	\hspace{-.001cm}\text{s.t.}:\,  \eqref{Opti_prob1}, \nonumber
 	\\& \label{C_rho}\hspace{.5cm} (2-\int\limits_{}^{} {{\xi _0}{\left( y \right)}} f_j^{*} dy -\int\limits_{}^{} {{\xi _1}{\left( w \right)}} f_w^{*} dw )\hspace{-.1cm}\mid_{\vartheta^*}\ge 1-\varepsilon.
 	\end{align}
 \end{subequations}
 As the problem has an optimization variable $\rho$, first we obtain the feasible set of $\rho$ which satisfies constraint \eqref{C_rho} through numerical methods. Next, we solve the following convex optimization problem with available software such as the CVX solver \cite{CVX}, 
 
  	\begin{subequations}\label{MethodI_rho_final}
 	\begin{align}
 	&\min_{\rho}\; {\rho ^{ - 1}} + {\rho ^{ - 1}}{\gamma _b},\label{}\\& 
 	\text{s.t.}:  \eqref{Opti_prob1}, \nonumber
 	\\&  \hspace{.18cm}\label{C_rh}\hspace{.36cm}  \rho \in \mathcal{F},
 	\end{align}
 \end{subequations}
where $\mathcal{F}$ is the feasible set of the optimization variable $\rho$ which satisfies constraint \eqref{C_rho}.

 Finally, in order to solve optimization problem \eqref{Opt_method1}, we propose an iterative algorithm which is given in Algorithm \ref{Alg_M1}.
  \begin{algorithm}[t] 
 	\caption{PROPOSED ITERATIVE POWER ALLOCATION ALGORITHM FOR SCENARIO I} \label{Alg_M1}
 	\begin{algorithmic}[1]
 		\STATE  \nonumber
 	Initialization: Set $k =0$ ($k$ is the iteration number)
 	and initialize to the power allocation factor $\rho(0)$,
 	\STATE 
 	Set $\rho=\rho( k)$,
 	\STATE  		
 	Solve \eqref{min_nu} by employing Algorithm \ref{Alg_bs} and set  the outcome  to $\vartheta(k+1)$,
 	\STATE 
 	Solve \eqref{MethodI_rho_final}  and set  the outcome  to $\rho(k+1)$,
 	\STATE 
 	If $\left| {\rho (k + 1) - \rho (k)} \right| \le \tau $ and  $\left| {\vartheta (k + 1) - \vartheta (k)} \right| \le \tau $\\
 	Stop,\\
 	else\\
 	set $k = k + 1$ and go back to step \ref{set}.
 	\end{algorithmic}
 \end{algorithm}
 This algorithm is stopped when the stopping conditions $ \left| {\rho (k + 1) - \rho (k)} \right| \le \tau$ and  $\left| {\vartheta (k + 1) - \vartheta (k)} \right| \le \tau $ are satisfied, where $k$ and  $\tau$ are the iteration number and the stopping threshold, respectively.

   \subsection{MV Scenario}\label{M2}
In many situations, the estimation of some statistical information such as the mean and variance are practical and easier than estimation of the full channel distribution.  Define sets $\mathcal{K}_1$ and $\mathcal{K}_2$ to have all possible distributions with a given mean and variance.  In other words,  $ \mathcal{K}_1 = \left\{ {{f_a}|{\mu _{{h_a}}} = {\rm{E}}\left\{ {{h_a}} \right\},\sigma _{{h_a}}^2 = {\rm{E}}\left\{ {h_a^2} \right\} - \mu _{{h_a}}^2} \right\}, \, \mathcal{K}_2= \left\{ {{f_j}|{\mu _{{h_j}}} = {\rm{E}}\left\{ {{h_j}} \right\},\sigma _{{h_j}}^2 = {\rm{E}}\left\{ {h_j^2} \right\} - \mu _{{h_j}}^2} \right\}$, where $\rm E \left\{ . \right\}$ is the expectation  operator. In this scenario, our optimization problem is formulated as follows:
\begin{subequations}\label{MethodII}
  	\begin{align}
&\max_{\rho}\;\mathbb{P}_{\psi_1}  \log \left( {1 + \frac{{\rho {\gamma _b}}}{{1 + \left( {1 - \rho } \right){\gamma _b}}}} \right),\label{MethodII0}\\& 
\hspace{.15cm}\text{s.t.}: \hspace{.18cm}  \eqref{Opti_prob1}, \nonumber
\\&\hspace{.85cm}  \min_{\vartheta} \mathop {\inf }\limits_{\begin{array}{*{20}{l}}
	{_{f_a \in \mathcal{K}_1}}\\
	{_{f_j \in \mathcal{K}_2}}
	\end{array}} \left( {{\mathbb{P}_{FA}} + {\mathbb{P}_{MD}}} \right) \ge 1 - \varepsilon. \label{MethodII2}
\end{align}
\end{subequations}
Constraint \eqref{MethodII2} is not in closed-form; hence, it is difficult to solve the optimization problem \eqref{MethodII}. To tackle this issue, we again are pessimistic from Alice's perspective and assume a lower bound for  \eqref{MethodII2}  to obtain a constraint that leads to a tractable problem. In particular, we find  lower bounds for ${\mathbb{P}_{FA}}$ and ${\mathbb{P}_{MD}}$  through the use of probabilistic inequalities. In order to obtain these lower bounds for ${\mathbb{P}_{MD}}$ and ${\mathbb{P}_{FA}}$,
we employ Markov's inequality and Cantelli's inequality, respectively, \cite{D. P. Dubhashi},  \cite{ME}. For non-negative random variable $X$  with mean $\mu$, $\forall a>0,$ Markov's inequality states:
\begin{align}\label{Makoves_ineq}
{\mathbb{P}(X \ge a) \le \frac{\mu}{a}},
\end{align}
 Moreover, Cantelli's inequality states that for real-valued random variable $X$ with mean $\mu$ and variance $\sigma^2$, the following inequity is satisfied  \cite{D. P. Dubhashi}:

\begin{align}\label{Centalli_ineq}
\mathbb{P}(X - \mu  \ge t) \left\{ {\begin{array}{*{20}{l}}
	{ \le \frac{{{\sigma ^2}}}{{{\sigma ^2} + {t^2}}}}&{t > 0,}\\
	{ \ge 1 - \frac{{{\sigma ^2}}}{{{\sigma ^2} + {t^2}}}}&{t < 0.}
	\end{array}} \right.
\end{align}
By employing \eqref{Makoves_ineq},  we have:

\begin{align}\label{inf_pmd}
&{\mathbb{P}_{MD}}  =\mathbb{P} \left( {\sigma _w^2 +\frac{{\rho {P}{{\left| {{h_{aw}}} \right|}^2}}}{{d_{aw}^\alpha }} + \frac{{\left( {1 - \rho } \right){P}{{\left| {{h_{jw}}} \right|}^2}}}{{d_{jw}^\alpha }} < \vartheta } \right), \nonumber\\& =1- \mathbb{P} \left( {\sigma _w^2 + \frac{{\rho {P}{{\left| {{h_{aw}}} \right|}^2}}}{{d_{aw}^\alpha }} + \frac{{\left( {1 - \rho } \right){P}{{\left| {{h_{jw}}} \right|}^2}}}{{d_{jw}^\alpha }} \ge \vartheta } \right),\nonumber\\&\ge 1 - \frac{{\sigma _w^2 + \frac{{\rho {P}{\mu _{{h_a}}}}}{{d_{aw}^\alpha }} + \frac{{\left( {1 - \rho } \right){P}{\mu _{{h_j}}}}}{{d_{jw}^\alpha }}}}{\vartheta }.
\end{align}
By employing  \eqref{Centalli_ineq}, the lower bound of ${\mathbb{P}_{FA}}$ is written as follows:
\begin{align}\label{inf_pfa}
{\mathbb{P}_{FA}}& =\mathbb{P} \left( {\sigma _w^2 + \frac{{{{\left| {{h_{jw}}} \right|}^2}}}{{d_{jw}^\alpha }}{\left( {1 - \rho } \right)P} > \vartheta } \right) \nonumber \\&=\mathbb{P} \left( {\sigma _w^2 + \frac{{{{\left| {{h_{jw}}} \right|}^2}}}{{d_{jw}^\alpha }}{\left( {1 - \rho } \right)P} -\mu_{fa} > \vartheta -\mu_{fa} } \right)\nonumber
\\& \ge 1 - \frac{\sigma_{fa} ^2}{\sigma_{fa} ^2 + \left( {\vartheta  - {\mu _{fa}}} \right)^2}, \,\, \forall \vartheta\le  \mu_{fa}, 
\end{align}
where $\mu_{fa}={\sigma _w^2 + \frac{{{\mu _{{h_j}}}}}{{d_{jw}^\alpha }}{\left( {1 - \rho } \right)P}}$ and $\sigma_{fa} ^2={\frac{{\sigma _{{h_j}}^2}}{{d_{jw}^{2\alpha }}}\left(\left( {1 - \rho } \right)P\right)^2}$.
 By employing \eqref{inf_pmd} and \eqref{inf_pfa}, the inner $\inf$ in \eqref{MethodII2} can be calculated as follows:
\begin{align}\label{inf}
&\mathop {\inf }\limits_{\begin{array}{*{20}{l}}
	{_{f_a \in \mathcal{K}_1}},‌\,
	{_{f_j \in \mathcal{K}_2}}
	\end{array}} \left( {{\mathbb{P}_{FA}} + {\mathbb{P}_{MD}}} \right)= 2-\\& \frac{\sigma_{fa} ^2}{\sigma_{fa} ^2 + \left( {\vartheta  - {\mu _{fa}}} \right)^2}- \frac{{\sigma _w^2 + \frac{{\rho {P}{\mu _{{h_a}}}}}{{d_{aw}^\alpha }} + \frac{{\left( {1 - \rho } \right){P}{\mu _{{h_j}}}}}{{d_{jw}^\alpha }}}}{\vartheta }, \, \forall \vartheta\le  \mu_{fa} \nonumber
\end{align}
 By using \eqref{inf}, \eqref{MethodII} is reformulated as:
\begin{subequations}\label{opt_methodII}
	\begin{align}
	&\max_{\rho}\;\mathbb{P}_{\psi_1}  \log \left( {1 + \frac{{\rho {\gamma _b}}}{{1 + \left( {1 - \rho } \right){\gamma _j}}}} \right),\label{opt_methodII0}\\& 
	\hspace{.15cm}\text{s.t.}: \hspace{.18cm}  \eqref{Opti_prob1}, \nonumber
	\\&\hspace{.85cm}  \min_{\vartheta} \left\{2- {\frac{{\sigma _w^2 + \frac{{\rho {P}{\mu _{{h_a}}}}}{{d_{aw}^\alpha }} + \frac{{\left( {1 - \rho } \right){P}{\mu _{{h_j}}}}}{{d_{jw}^\alpha }}}}{\vartheta }} \right.  \label{opt_methodII2}\\&\hspace{.85cm}  -\left. {\frac{\sigma_{fa} ^2}{\sigma_{fa} ^2 + \left( {\vartheta  - {\mu _{fa}}} \right)^2}} \right\}
	 \ge 1 - \varepsilon, \,\, \forall \vartheta\le  \mu_{fa}.  \nonumber
	\end{align}
\end{subequations}
The disadvantage of this approach is that the Cantelli inequality is usable subject to $ \vartheta\le  \mu_{fa}$. For solving the optimization problem \eqref{opt_methodII}, we exploit the well-known iterative algorithm  called Alternative Search Method (ASM) \cite{K.Son}, in which the  problem is converted to two subproblems:  which one of them employs $\vartheta$ as the optimization variable, and the other employs $\rho$.  In each iteration, we find optimal $\vartheta$ and  $\rho$ separately. In other words,  we find optimal $\vartheta$ by considering fixed $\rho$  and vice versa.  This is shown in Algorithm \ref{Alg1}.  The algorithm is stopped when the stopping condition $ \left| {\rho (k + 1) - \rho (k)} \right| \le \tau$ is satisfied, where $\tau$ is the stopping threshold and $k$ is the iteration number. The optimization variable $\vartheta$ only exists in constraint \eqref{opt_methodII2}. Hence, the optimal  $\vartheta$ can be obtained by solving \eqref{opt_methodII2}, and we employ the Geometric programing (GP) method for solving it. Therefore, the left side of \eqref{opt_methodII2} is equivalent to the following problem in the GP format:
	\begin{subequations}\label{opt_methodII_varthete}
		\begin{align}
		&\min_{\vartheta,\, t, t_1, t_2}\;t^{-1},\label{}
		\\& \hspace{.15cm}\text{s.t.}: \hspace{.18cm}   tt_3^{ - 1} \le 1,
		\\&\hspace{.85cm} \sigma _{fa}^2t_1^{ - 1}t_3^{ - 1} + \sigma _w^2{\vartheta ^{-1}}t_3^{ - 1} +\\&\hspace{.85cm} \left( {\frac{{\rho {P}{\mu _{{h_a}}}}}{{d_{aw}^\alpha }} + \frac{{\left( {1 - \rho } \right){P}{\mu _{{h_j}}}}}{{d_{jw}^\alpha }}} \right){\vartheta ^{-1}}t_3^{ - 1} \le 1, \nonumber
		\\&\hspace{.85cm} \sigma _{fa}^2t_2^{ - 1} + {\mu _{fa}^2}t_2^{ - 1} + {\vartheta ^2}t_2^{ - 1} \le 1,
		\\&\hspace{.85cm} {t_1}t_2^{ - 1} + t{\mu _{fa}}\vartheta t_2^{ - 1} \le  1.
		\end{align}
	\end{subequations}
	For solving the  optimization problem \eqref{opt_methodII_varthete}, we can employ available softwares such as the CVX solver \cite{CVX}.

After obtaining the optimal $\vartheta$,  we aim to obtain the optimal $\rho$ with fixed $\vartheta$. By the same argument which is employed in Section \ref{M1},  we are able to replace  $\min_{\rho}\; {\rho ^{ - 1}} + \left(\rho ^{ - 1}-1\right){\gamma _j}$ with  $\max_{\rho}\;\mathbb{P}_{\psi_1}  \log \left( {1 + \frac{{\rho {\gamma _b}}}{{1 + \left( {1 - \rho } \right){\gamma _j}}}} \right)$.
 In the following, we employ the GP method to solve \eqref{opt_methodII}. Therefore, the optimization problem \eqref{opt_methodII} is equivalent to the following problem with the GP format:
	\begin{subequations}\label{opt_methodII_rho}
		\begin{align}
		&\min_{\rho}\; {\rho ^{ - 1}} + {\rho ^{ - 1}}{\gamma _j},\label{}
		\\&\hspace{.15cm}\text{s.t.}: \hspace{.18cm}  \eqref{Opti_prob1}, \nonumber
		\\&\hspace{.85cm}{\vartheta ^{ - 1}}\left( {\frac{{\sigma _w^2 + \rho {P}\mu _{{h_a}}}}{{d_{aw}^\alpha }} + \frac{{{P}\mu _{{h_j}}}}{{d_{jw}^\alpha }}} \right) + {q^{ - 1}} \le {q_5},
		\\&\hspace{.85cm}1 + \varepsilon  + \left( {\frac{{\rho {P}\mu _{hj}^{}}}{{d_{jw}^\alpha }}} \right){\vartheta ^{ - 1}} \le {q_5},
		\\&\hspace{.85cm} \left( {{\rho ^2} + 1} \right)P^2\sigma _{{h_j}}^2q_2^{ - 1} \le 1,
		\\&\hspace{.85cm} {q_1}q_2^{ - 1} + 2\rho P^2\sigma _{{h_j}}^2q_2^{ - 1} \le 1, 
		\\&\hspace{.85cm} q_4^{ - 1} + {\left( {\vartheta d_{jw}^\alpha  - {\mu _{{h_j}}}{P} - \sigma _w^2d_{jw}^\alpha } \right)^2}q_1^{ - 1}q_4^{ - 1}+\\&\hspace{.85cm}  {\rho ^2}\mu _{{h_j}}^2P^2q_1^{ - 1}q_4^{ - 1} + 2\rho {\mu _h}_j{P}\vartheta d_{jw}^\alpha q_1^{ - 1}q_4^{ - 1} \le 1,
		\nonumber 
			\\&\hspace{.85cm} 2\rho {\mu _{{h_j}}}{P}\left( {{\mu _{{h_j}}}{P} + \sigma _w^2d_{jw}^\alpha } \right)q_1^{ - 1}q_4^{ - 1} + qq_4^{ - 1} \le 1,
		\end{align}
	\end{subequations}
	For solving the  optimization problem \eqref{opt_methodII_rho}, we can again employ available softwares such as the CVX solver \cite{CVX}.
\begin{algorithm}[t] 
	\caption{PROPOSED ITERATIVE POWER ALLOCATION ALGORITHM FOR SCENARIO II} \label{Alg1}
	\begin{algorithmic}[1]
		\STATE  \nonumber
		Initialization: Set $k =0$ and initialize to the power allocation factor $\rho(0)$,
		\STATE \label{set}
		Set $\rho=\rho( k)$,
		\STATE  		
		Solve \eqref{opt_methodII_varthete} and set  the outcome  to $\vartheta(k+1)$,
		\STATE 
		Solve \eqref{opt_methodII_rho}  and set  the outcome  to $\rho(k+1)$,
		\STATE 
		If $\left| {\rho (k + 1) - \rho (k)} \right| \le \tau $,\\
		stop,\\
		else\\
		set $k = k + 1$ and go back to step \ref{set}.
	\end{algorithmic}
\end{algorithm}	
	
\subsection{FCDI-UV Scenario} \label{M3}
	\textcolor{black}{
In this section, we investigate a situation in which  legitimate transmitters  know Willie's channel distribution is complex Gaussian but the variance is unknown. The channel coefficients for the channels from Alice to Willie and the jammer to Willie are complex Gaussian with zero mean and variance $\sigma_a^2$ and $ \sigma_j^2$, respectively, i.e., ${h_{aw}} \sim \mathcal{CN}\left( {0, \sigma_a^2} \right)$ and ${h_{jw}} \sim \mathcal{CN}\left( {0, \sigma_j^2} \right)$,  where $\sigma_a^2$ and $\sigma_j^2$ are unknown at the legitimate transmitters. Hence, 
 $\left| {{h_{aw}}} \right|\sim \text{Rayleigh}\left( {\sqrt {\sigma_a^2} } \right)$, $\left| {{h_{jw}}} \right|\sim \text{Rayleigh}\left( {\sqrt {\sigma_j^2} } \right)$,   ${\left| {{h_{aw}}} \right|^2}\sim \text{exponential}\left( {\sigma_a^2} \right)$, and ${\left| {{h_{jw}}} \right|^2}\sim \text{exponential}\left( {\sigma_j^2} \right)$.  Then, the pdfs of the random variables in \eqref{gamma} are given by:
\begin{align}\label{distribution}
{f_{{\Gamma _\Psi}}}\left( \gamma  \right) = \left\{ {\begin{array}{*{20}{l}}
	{\frac{1}{{{\psi _0}}}{e^{ - \frac{\gamma }{{{\psi _0}}}}},}&{{\rm{ }}{\Psi _0},}\\
	{\frac{1}{{{\psi _1} - {\psi _0}}}\left[ {{e^{ - \frac{\gamma }{{{\psi _1}}}}} - {e^{ - \frac{\gamma }{{{\psi _0}}}}}} \right],}&{{\rm{}}{\Psi _1},}
	\end{array}} \right.
\end{align}
where $\psi_0= \frac{\left(1-\rho \right)P \sigma_j^2}{d_{jw}^\alpha}$ and $\psi_1= \frac{\rho P \sigma_a^2}{d_{aw}^\alpha}$.  By employing \eqref{distribution}, we have:
\begin{align}
\begin{array}{*{20}{l}}
{}&{\mathbb{P}_{FA} = \left\{ {\begin{array}{*{20}{l}}
		{{e^{ - \frac{1}{{{\psi _0}}}\left( {\vartheta  - \sigma _w^2} \right)}},}&{\vartheta  - \sigma _w^2 > 0,}\\
		{}&{}\\
		{1,}&{\vartheta  - \sigma _w^2 \le 0.}
		\end{array}} \right.}\\
{}&{\mathbb{P}_{MD} = }\\
&\hspace{-.5cm}{\left\{ {\begin{array}{*{20}{l}}
		{\begin{array}{*{20}{l}}
			{1 + \frac{1}{{{\psi _1} - {\psi _0}}}}\\
			{ \times \left[ { - {\psi _1}{e^{ - \frac{1}{{{\psi _1}}}\left( {\vartheta  - \sigma _w^2} \right)}} + {\psi _0}{e^{ - \frac{1}{{{\psi _0}}}\left( {\vartheta  - \sigma _w^2} \right)}}} \right]}
			\end{array}}&{\vartheta  - \sigma _w^2 > 0,}\\
		{\begin{array}{*{20}{l}}
			{}
			\end{array}}&{}\\
		0&{\vartheta  - \sigma _w^2 \le 0.}
		\end{array}} \right.{\rm{ }}}
\end{array}
\end{align}
Willie selects $\vartheta > \sigma _w^2$, because otherwise ${\mathbb{P}_{FA}} + {\mathbb{P}_{MD}}=1$. For mathematical simplification, we assume $\sigma_a^2=\sigma_j^2=\sigma$}. 	\textcolor{black}{Please note that this assumption is reasonable because the nodes are located in the same environment; hence, although the variances are unknown, it is reasonable to assume that the channel coefficients have the same variance, i.e., $\sigma$.}	\textcolor{black}{
In this scenario, we again are pessimistic from Alice's perspective and investigate the worst-case scenario; therefore,  our optimization problem is formulated as:
\begin{subequations}\label{MethodIII}
	\begin{align}
	&\max_{\rho}\;\mathbb{P}_{\psi_1}  \log \left( {1 + \frac{{\rho {\gamma _b}}}{{1 + \left( {1 - \rho } \right){\gamma _j}}}} \right),\label{MethodIII0}\\& 
	\hspace{.15cm}\text{s.t.}: \hspace{.18cm}  \eqref{Opti_prob1}, \nonumber
	\\&\hspace{.85cm}  \min_{\vartheta} \min_{\sigma}  {e^{ - \frac{1}{{{\psi _0}}}\left( {\vartheta  - \sigma _w^2} \right)}} +1 + \frac{1}{{{\psi _1} - {\psi _0}}} \nonumber\\& \hspace{.85cm} \times \left[ { - {\psi _1}{e^{ - \frac{1}{{{\psi _1}}}\left( {\vartheta  - \sigma _w^2} \right)}} + {\psi _0}{e^{ - \frac{1}{{{\psi _0}}}\left( {\vartheta  - \sigma _w^2} \right)}}} \right]  \ge 1 - \varepsilon. \label{MethodIII2}
	\end{align}
\end{subequations}
We first solve the inner minimization in \eqref{MethodIII2}.
Taking the derivative of the left side of \eqref{MethodIII2} with respect to $\sigma$, we have 
\begin{align}
&\frac{{d\left( {{\mathbb{P}_{FA}} + {\mathbb{P}_{MD}}} \right)}}{{d\sigma}} =\\& \frac{{\left( {\rho {\mkern 1mu} d_{jw}^\alpha {{\rm{e}}^{\frac{{d_{jw}^\alpha \left( { - {\sigma _w^2} + \vartheta } \right)}}{{\left( { - 1 + \rho } \right){P}\sigma}}}} + d_{aw}^\alpha {{\rm{e}}^{ - \frac{{d_{aw}^\alpha \left( { - \sigma _w^2 + \vartheta } \right)}}{{\rho {\kern 1pt} {P}\sigma}}}}\left( { - 1 + \rho } \right)} \right) }}{{\sigma ^2{P}{{\left( { - 1 + \rho } \right)}}\left( {\left( { - 1 + \rho } \right)d_{aw}^\alpha  + \rho d_{jw}^\alpha } \right)}}\times \nonumber\\&\left( { \sigma _w^2 - \vartheta} \right)d_{jw}^\alpha= 0. \nonumber
\end{align}
After some mathematical manipulation,  the optimal $\sigma$ is given by
\begin{align}
 {\sigma^*} = \frac{{\left( {\vartheta  - \sigma _w^2} \right)\left( {d_{aw}^\alpha \rho  + \rho {\mkern 1mu} d_{jw}^\alpha  - d_{aw}^\alpha } \right)}}{{\rho \ln \left( { - \frac{{\left( {\rho  - 1} \right)d_{aw}^\alpha }}{{d_{jw}^\alpha \rho {\mkern 1mu} }}} \right)\left( {\rho  - 1} \right){P}}}
 \end{align}
 Finally, the left side of \eqref{MethodIII2} is equivalent to \eqref{min_u_nu} at the top of the next page.
\begin{figure*}[h]    	 
\begin{align}\label{min_u_nu}
&\min_{\vartheta} \mathop {\min }\limits_{u} \left( {{\mathbb{P}_{FA}} + {\mathbb{P}_{MD}}} \right)  =1 + \frac{{\left( {\rho {\mkern 1mu} d_{jw}^\alpha {{\rm{e}}^{\frac{{\rho {\kern 1pt} \ln \left( { - \frac{{\left( { - 1 + \rho } \right)d_{aw}^\alpha d_{jw}^{ - \alpha }}}{{\rho {\kern 1pt} }}} \right)d_{jw}^\alpha }}{{\left( {\left( { - 1 + \rho } \right)d_{aw}^\alpha  + \rho {\kern 1pt} d_{jw}^\alpha } \right)}}}} - \rho {\mkern 1mu} d_{jw}^\alpha {{\rm{e}}^{ - \frac{{\ln \left( { - \frac{{\left( { - 1 + \rho } \right)d_{aw}^\alpha d_{jw}^{ - \alpha }}}{{\rho {\kern 1pt} }}} \right)\left( { - 1 + \rho } \right)d_{aw}^\alpha }}{{\left( {\left( { - 1 + \rho } \right)d_{aw}^\alpha  + \rho {\kern 1pt}d_{jw}^\alpha } \right)}}}}} \right)}}{{\left( {\left( { - 1 + \rho } \right)d_{aw}^\alpha  + \rho {\mkern 1mu} d_{jw}^\alpha } \right)}}
\end{align}
	\hrule
\end{figure*}
   Hence, the optimization problem  \eqref{MethodIII} can be rewritten as:
\begin{subequations}\label{MethodIII_final0}
	\begin{align}
	&\max_{\rho}\;\mathbb{P}_{\psi_1}  \log \left( {1 + \frac{{\rho {\gamma _b}}}{{1 + \left( {1 - \rho } \right){\gamma _b}}}} \right),\label{}\\& 
	\hspace{.15cm}\text{s.t.}: \hspace{.18cm}  \eqref{Opti_prob1} \nonumber\\& \hspace{.85cm} \frac{{\left( {1 - \rho } \right)d_{aw}^\alpha }}{{\rho d_{jw}^\alpha  - \left( {1 - \rho } \right)d_{aw}^\alpha }}\ln \left( {\frac{{\left( {1 - \rho } \right)d_{aw}^\alpha }}{{\rho d_{jw}^\alpha }}} \right) \le \ln \left(\varepsilon\right),
	\end{align}
\end{subequations} 
By utilizing an auxiliary variable $t$, the optimization problem \eqref{MethodIII_final0} is equivalent to the following problem
\begin{subequations}
	\begin{align}
&\max_{\rho}\;\mathbb{P}_{\psi_1}  \log \left( {1 + \frac{{\rho {\gamma _b}}}{{1 + \left( {1 - \rho } \right){\gamma _b}}}} \right),\label{}\\& 
\hspace{.15cm}\text{s.t.}: \hspace{.18cm}  \eqref{Opti_prob1} \nonumber\\& \label{m3_fb} \hspace{.85cm}\ln \left( {\frac{{\left( {1 - \rho } \right)d_{aw}^\alpha }}{{\rho d_{jw}^\alpha }}} \right) \times \left( {1 - \rho } \right)d_{aw}^\alpha  - t\ln \left( \varepsilon  \right) < 0,\\& \hspace{.85cm} \label{m3_fc}\rho d_{jw}^\alpha  - \left( {1 - \rho } \right)d_{aw}^\alpha  \le t
\end{align}
\end{subequations}
Moreover, by the same argument which is employed in Section \ref{M1},  we are able to replace $\min_{\rho}\; {\rho ^{ - 1}} + \left(\rho ^{ - 1}-1\right){\gamma _j}$ with  $\max_{\rho}\;\mathbb{P}_{\psi_1}  \log \left( {1 + \frac{{\rho {\gamma _b}}}{{1 + \left( {1 - \rho } \right){\gamma _j}}}} \right)$.  Finally, we have a convex optimization problem as follows:
	\begin{align}\label{MethodIII_final1}
	&\min_{\rho}\; {\rho ^{ - 1}} + {\rho ^{ - 1}}{\gamma _j}-\gamma_j,\\& 
	\hspace{.15cm}\text{s.t.}: \hspace{.18cm}  \eqref{Opti_prob1}, \eqref{m3_fb}, \eqref{m3_fc} \nonumber
	\end{align}
In order to solve the convex optimization problem \eqref{MethodIII_final1}, we can employ available softwares such as the CVX solver \cite{CVX}.}

\section{Scenario II: CDI is Known at Willie }\label{CKW}
In this scenario, we assume that  the CDI is known at Willie but still unknown at the legitimate transmitters. In this case, since Willie knows the CDI perfectly, he can calculate the threshold for his decision exactly based on the known channel distribution.
 But, since  Alice does not know the CDI, we first should find $\vartheta$ over all $f$s, and then finds the worst $f$. In other words, in this scenario,  
  \eqref{Opti_prob2} is replaced with $\mathop {\min }\limits_{\begin{array}{*{20}{l}}
 	_{f_w} , _{f_j}
 	\end{array}}\min\limits_{\vartheta}\left(\;{\mathbb{P}_{FA}} + {\mathbb{P}_{MD}}\right)\ge 1-\varepsilon, $ i.e., the optimization problem is as follows
 \begin{subequations}\label{ss}
 	\begin{align}
 	&\max_{\rho}\;\mathbb{P}_{\psi_1}  \log \left( {1 + \frac{{\rho {\gamma _b}}}{{1 + \left( {1 - \rho } \right){\gamma _j}}}} \right),\\& 
 	\hspace{.01cm} \text{s.t.}:\hspace{.01cm} \eqref{Opti_prob1} \nonumber\\&\label{ss2}\hspace{.05cm} 
 	\hspace{.008cm}\mathop {\min }\limits_{\begin{array}{*{20}{l}}
 		_{f_w} , _{f_j}
 		\end{array}}\hspace{-.2cm}\min\limits_{\vartheta}\left(\int\limits_{\frac{{\left( {\vartheta  - \sigma _w^2} \right)d_{jw}^\alpha }}{{\left( {1 - \rho } \right){P}}}}^\infty  {{f_j}\,d{y}}  + \int\limits_0^{\vartheta  - \sigma _w^2} {{f_w}\,dw} \right)\ge 1-\varepsilon,
 	\end{align}
 \end{subequations}
In order to solve optimization problem \eqref{ss}, 
first we investigate \eqref{ss2}. The inner minimum on the left side of \eqref{ss2} is solved with a Particle Swarm Optimization (PSO) algorithm \cite{J.Kennedy}, \cite{R.Eberhart} by employing Algorithm \ref{Alg4}. Next, we find the worst case $f_w$ and $f_j$ with the mentioned policies in  \ref{M1}, \ref{M2}, and \ref{M3}.

\begin{algorithm}[t] 
	\caption{PSO for inner minimum in the left side of \eqref{ss2} } \label{Alg4}
	\begin{algorithmic}[1]
		\STATE  \nonumber
		\textbf{Initialization}: 
		
		\textbf{Step 1)}
	MT: Iteration number of PSO algorithm,

		\hspace{1.3cm}	nP: Number of particles of PSO algorithm,
		  		
	   \hspace{1.3cm}	For $\vartheta$: 
		
	\hspace{1.7cm}	$\Phi_i$: Position of one particle, $i = 1, 2, . . . ,$ nP
		
	\hspace{1.7cm}  $V_i$: Velocity of one particle, $i = 1, 2, . . . ,$ nP

		\textbf{Step 2)} Evaluate inner minimum in the left side of \eqref{ss2}
		 as a cost for all particles, named $C_i$

       \hspace{1.3cm}  Set $bestp_i=\Phi_i$ and $bestp_i.C_i=C_i$
       
        \hspace{1.3cm}  Set $bestg$ and $bestg.C$ equal to the best
        
      \hspace{1.4cm}  initial particle 
      
\hspace{-.7cm}\textbf{---------------------------------------------------------------------------}
       \STATE  \nonumber
       \textbf{Main Loop}: 
       
       $for\,\, q=1:MT$

       \hspace{.2cm}  $for\,\, i=1:nP$
       
       \hspace{.9cm}  Update the velocity and position of particles 
       
              \hspace{1.3cm}
       according to \cite{F.Alavi}
        
       \hspace{.9cm}         Evaluate inner minimum in the left side of \eqref{ss2}

       \hspace{.9cm}        if ($C_i < bestp.C_i$)
        
       \hspace{1.1cm}        do:  $bestp_i=\Phi_i$ and $bestp. C_i= C_i$ 
        
       \hspace{1cm}        if ($ bestp. C_i< bestg. C$)
        
       \hspace{1.2cm}        do: $bestg=bestp_i$ and $bestg.C=bestp.C_i$
        
               \hspace{.4cm}end 
        
        \hspace{.001cm}       end

	\end{algorithmic}
\end{algorithm}	

\section{Numerical Results}\label{Numerical Results}

In this section, we present numerical results to evaluate the performance of the proposed schemes in terms of covert rate in the case when the channel distribution information for the channels from the transmitters (Alice, jammer) to Willie is uncertain at Alice.   For comparison purposes, we will also consider the case of ``perfect CDI''; in this case, both Alice and Willie know the CDI exactly.

The simulation parameters employed are: the noise power at Willie is $\sigma_w^2=-10$ dBW,  the noise power at Bob is $\sigma_b^2= -10$ dBW,  the probability of data transmission is $\mathbb{P}_{\psi_1}=0.5$, and the covertness requirement is $\epsilon=0.1$.

  \begin{figure}[h!]
  		\includegraphics[width=3.8in,height=3in]{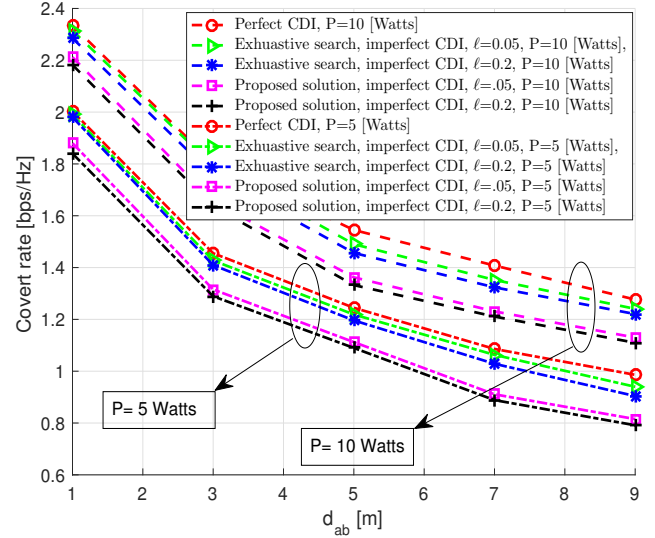}
 		\caption{Covert rate versus $d_{ab}$ when Willie knows CDI perfectly and legitimate transmitters know CDI imperfectly in the N-pdf scenario,  $d_{aw}=7$ m, $d_{jb}=7 $m, and $d_{jw}=5 $m.}
		\label{M1_dad_d}
  \end{figure}
Fig. \ref{M1_dad_d} depicts the covert rate versus the distance between Alice and Bob with the locations of the other nodes fixed when Willie knows the CDI perfectly and the legitimate transmitters (Alice, in particular) know CDI imperfectly in the N-pdf scenario. The curves are plotted for different values of the maximum distance $\ell$ around the nominal distribution and show the degree to which increasing $\ell$ causes the covert rate to decrease.  We also show the gap in performance between the solution found through our efficient search method and that found by exhaustive search.   In particular, the gap between the proposed solution and exhaustive search is roughly 8.8\%. 
Finally, we illustrate the loss in performance in the N-pdf scenario relative to the case where Alice has perfect CDI, which has been investigated in \cite{Forouzesh}.  As can be seen, there is approximately a 4.7\% gap between the case when Alice has perfect CDI and the N-pdf scenario.   This figure can also be employed to explore the impact of increased total transmit power, and it shows that by increasing total transmit power by a factor of two that the covert rate increases $1.2$ times.

\begin{table*}[t!] 
	\centering
	\caption{Comparison among methods, in points of view  efficiency and accuracy}
	\begin{tabular}{|m{8 em} |m{15 em}|m{19 em} |m{15 em} |} 
		\hline \rowcolor[gray]{0.8}
		Items & Features  &  Efficiency of the Proposed Solutions (Gap with exhaustive search) &  Performance Loss under Uncertain CDI (Gap with perfect CDI scenario) \\
		\hline
		MV scenario  & Availability of only mean and variance of channel distribution  & High, 1.47\% & High, 10.6\% \\ [0.5ex]  
		\hline
		N-pdf scenario &  Availability of nominal probability density function of channel & Medium, 8.8\%  & Medium, 4.7\% \\
		\hline
		FCDI-UV scenario & Availability of distribution with unknown variance   & Low, 31.6\% & Low,  1.7\%\\
		\hline
	\end{tabular}
	\label{table}
\end{table*}

\begin{figure}[t]
	\includegraphics[width=3.8in,height=3in]{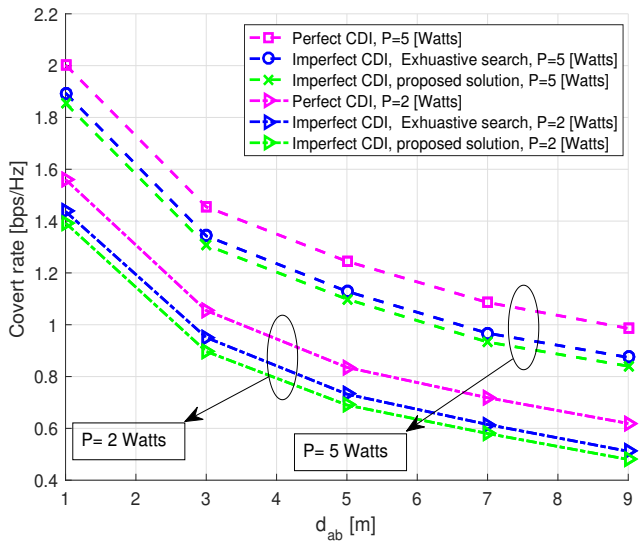}
	\caption{Covert rate versus $d_{ab}$ when Willie knows CDI perfectly and legitimate transmitters know CDI imperfectly  in the MV scenario, $d_{aw}=7$ m, $d_{jb}=7 $m,  $d_{jw}=5 $m, and $P= 5$ Watts.}
	\label{M2_dab_pmax}
\end{figure}

In Fig. \ref{M2_dab_pmax}, the  covert rate versus $d_{ab}$  when Willie knows the CDI perfectly and legitimate transmitters know CDI imperfectly in the MV scenario is shown.  As can be seen, there is a 10.6\% gap between the perfect CDI and the MV scenario, approximately. Furthermore, this figure evaluates the performance of the proposed optimization problem solution by comparing it with exhaustive search, and we see that there is a 4\% performance loss relative to exhaustive search.  Moreover, it shows that by increasing total transmit power two times the covert rate increases $1.2$ times. 

\begin{figure}[t]
	\includegraphics[width=3.8in,height=3in]{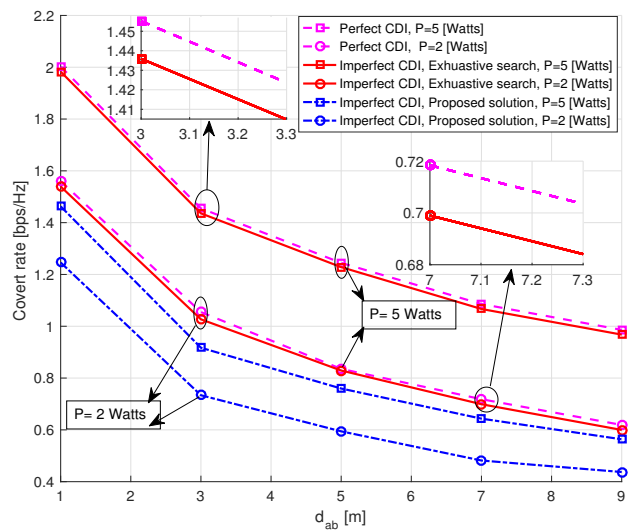}
	\caption{Covert rate versus $d_{ab}$ when Willie knows CDI perfectly and legitimate transmitters know CDI imperfectly  in the FCDI-UV scenario, $d_{aw}=7$ m, $d_{jb}=7 $m, and $d_{jw}=5 $m.}
	\label{M3_dab_pmax}
\end{figure}

\begin{figure}[t]
	\begin{center}$
		\begin{array}{cc}
\hspace{-0.45cm}	\subfigure[h][N-pdf scenario.]{
	\includegraphics[width=2 in,height=1.9in]{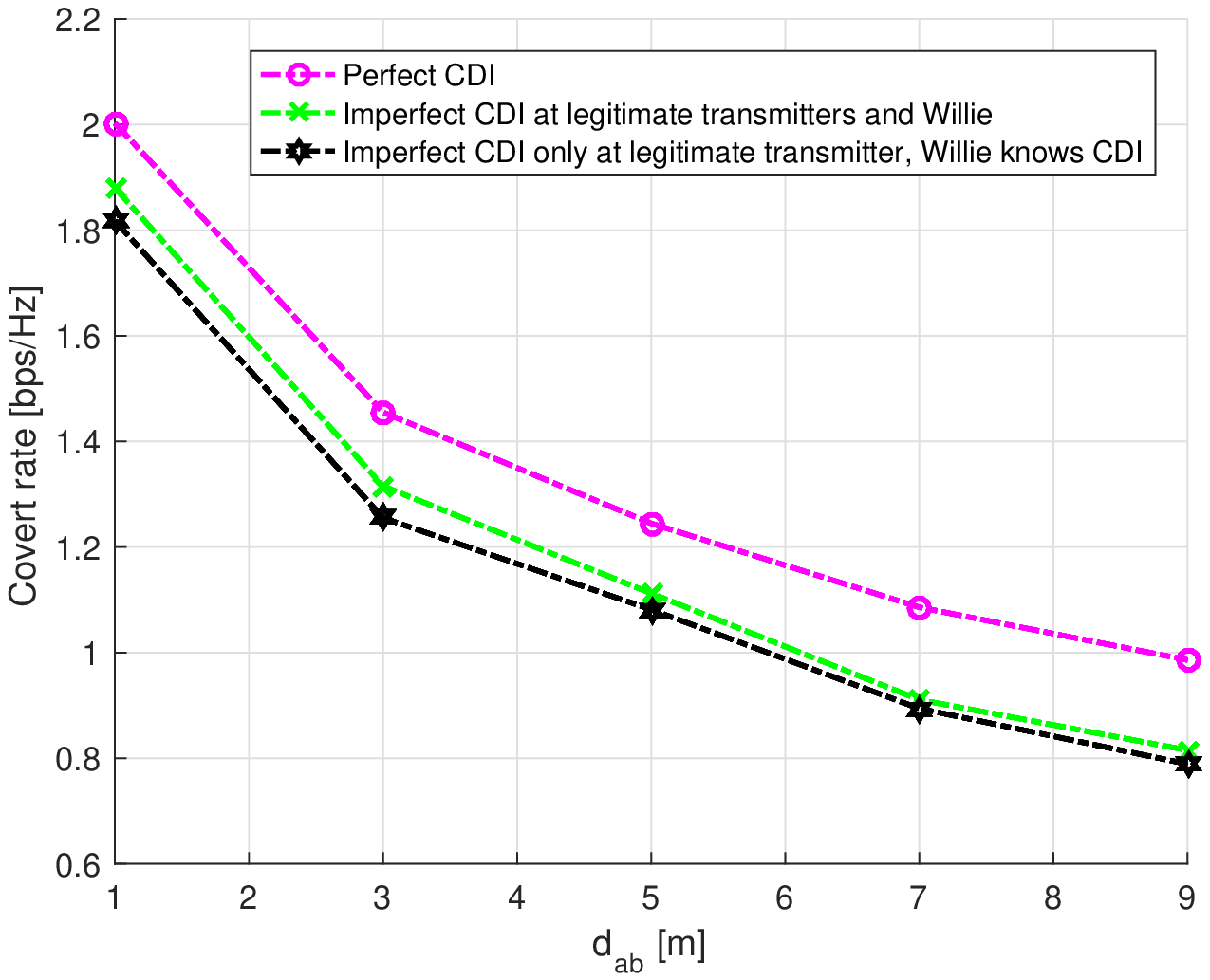}
	\label{SCW1}}
	\hspace{-0.69cm}
	\subfigure[h][MV scenario.]{
	\includegraphics[width=2 in,height=1.9in]{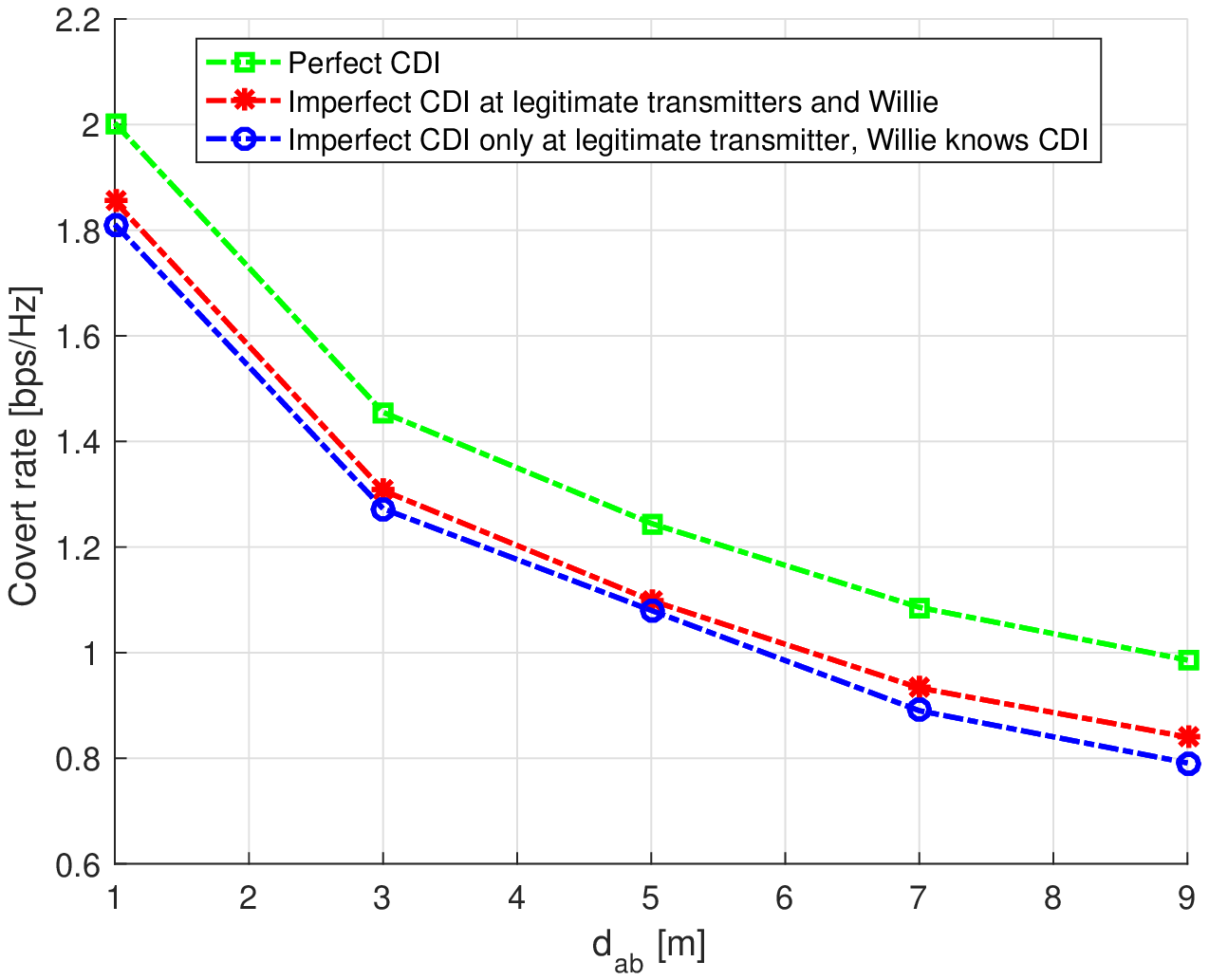}
	\label{SCW2}}
		\end{array}$
	\end{center}
	\caption{Comparison between scenarios of imperfect and perfect CDI at Willie, $d_{aw}=7$ m, $d_{jb}=7 $m,  $d_{jw}=5 $m , and $P=5$ Watts.}
\label{SCW}
\end{figure}
Fig. \ref{M3_dab_pmax} evaluates the optimal covert rate versus $d_{ab}$ when Willie knows CDI perfectly and legitimate transmitters know CDI imperfectly  in the FCDI-UV scenario.
The curves are depicted for different values of total transmit power. Imperfect CDI  in the FCDI-UV scenario leads to a 1.7\% gap in performance relative to the case when Alice has perfect CDI. َAs seen in this figure, the gap between exhaustive search and proposed solution is  31.6\%.

Fig. \ref{SCW} evaluates the impact of the availability of CDI at Willie. Fig. \ref{SCW}(a) evaluates this issue when the nominal probability density function of channels are available at legitimate transmitters (N-pdf), and  Fig.  \ref{SCW}(b) evaluates it when only the mean and variance of the channel distribution are available at legitimate transmitters (MV).
As expected, perfect CDI at Willie decreases the covert rate compared to the case when he has imperfect CDI, but the degree to which it decreases significantly depends on the nature of the uncertainty in the CDI.

A summary of the efficiency of our solutions versus exhaustive search and the loss in covert rate when CDI is uncertain is given in Table \ref{table}. 

\section{Conclusion}\label{Conclusion}
In order to design covert communication schemes, a calculation of the detection error at the adversary warden Willie under different approaches is necessary, but it is often difficult for the transmitter Alice to have an accurate characterization of the channel between herself and Willie. Unlike prior studies which assume the transmitter precisely knows this channel distribution information (CDI), hence risking a loss of covertness when this knowledge is inaccurate, we assume uncertainty in Alice's knowledge of such.  In particular, we have considered covert communication under partial and uncertain channel distribution information in the presence of a single warden Willie, legitimate transmitter Alice, a jammer, and legitimate receiver Bob. To this end, we proposed schemes for each of three different characterizations of the uncertainty in the channel distribution: 1) when the transmitter knows a nominal probability density function, 2) when only the mean and variance of the channel distribution are available to the transmitter, 3) when the transmitter knows the channel distribution is complex Gaussian but the  variance is unknown. Moreover, we investigated two possible assumptions: 1) CDI is known at Willie, 2) CDI is unknown at Willie. We have then considered optimal power allocations that maximize covert rate subject to the covertness requirement in each case. The power allocation optimization problems in some scenarios are non-convex and intractable, and hence we have proposed iterative solutions. In order to investigate the optimality gap, the performance of the proposed solutions was compared to that of solutions obtained with an exhaustive search. Numerical results illustrate the accuracy of each of schemes and the optimality gap of the proposed solutions.

\end{document}